# Effect of magnetization process on levitation force between a superconducting disk and a permanent magnet

L. Liu, Y. Hou, C.Y. He, Z.X. Gao

*Department of Physics, State Key Laboratory for Artificial Microstructure and Mesoscopic Physics,*

*Peking University, Beijing 100871, P. R. China*

L. Xiao, H.T. Ren, Y.L. Jiao, M.H. Zheng

*General Research Institute for Non-ferrous Metals, Beijing 100088, China*

**Abstract**

The levitation forces between a permanent magnet and a coaxial superconducting disk after different magnetization processes are measured. Significant effect of the magnetization process on the levitation force is observed. Theoretical calculations of levitation force based on the critical state model with temperature and field-dependent critical current density, and the heat dissipation due to the flux motion are in perfect agreement with the experimental data.

## Introduction

It is well known that the very high trapped magnetic field in bulk high temperature superconducting disk (SD) and the levitation force between SD and a permanent magnet (PM) have possible industrial applications [1-7]. Several groups reported their theoretical and experimental results in recent years. Qin *et al*. reported the theoretical calculation of the hysteretic levitation force between a PM and a non-magnetized SD after zero-field-cooling (ZFC) from first principles [8]. Wang *et al.* considered the magnetization of the SD, and calculated the levitation force between a PM and a SD after field-cooling (FC) [9]. In the above calculations, however, the critical current density as a function of the magnetic field and temperature, as well as the heat dissipation in SD was not taken into account. Ohsaki *et al.* considered both factors and calculated the distributions of the current density and temperature in SD after a magnetic field pulse, but their calculations did not distinguish the normal state region from the superconducting state region in the pulsed field magnetization process [10]. Hou *et al.* proposed a systematic theoretic framework to discuss the basic equation of the current motion in SD by taking into account the heat dissipation, the temperature and field-dependent critical current density, and the different current dependent electric fields $E(j)$ in the normal state region and the superconducting state region [11]. According to this framework they simulated the trapped magnetic field in SD, which is in good agreement with experimental results of several groups [12-13]. However, no experiment results on levitation force are reported so far. In this paper, the levitation forces between a PM and a coaxial SD were measured after different magnetization processes. The rate $\beta$ at which the applied magnetic field is swept down in FC process was changed and significant effect on the levitation force was observed. The calculated levitation forces according to the systematic theoretic framework [11] coincide with the experimental data very well.

**Experimental**

The SD used in experiments is a melt-textured-grown (MTG) YBCO sample of 30 mm in diameter, 20 mm in thickness [14]. The permanent magnet (Nd-Fe-B) is the same size and the magnetic field on its surface is 0.53T. The SD sample is set in a container first. In ZFC process the SD is cooled by liquid nitrogen to 77 K below its critical temperature (92K) without applied magnetic field. In FC case, a chosen magnetic field (0.5T, 0.375T, 0.25T generated by the electromagnets) is applied on the SD along the axial direction of the disk at room temperature, then the SD is cooled below its critical temperature in liquid nitrogen and the applied magnetic field is swept down at a given rate $\beta$ (0.1T/s or 0.0025T/s) to zero.

After above process, we wait for 1 minute and set the SD on the sample holder in the levitation force measurement system, which is shown in Fig. 1. With a jack the SD can approach the PM the levitation forces in this process are measured by a force sensor, and the distance between the SD and the PM is measured precisely by a displacement sensor. These electrical signals are transmitted to the computer and processed by LABVIEW program.

## Theoretical

## Basic equations for current motion

According to the systematic theoretic framework [11] we consider the SD with radius $a$ and thickness $b$. The desired equation of motion for the current density $J(\mathbf{r},t)$ in the superconducting disk as following [9]:

$$\dot{J}(\mathbf{r},t) = \mu_0^{-1} \int_0^a d\rho' \int_0^b dz' Q_{cyl}^{-1}(\mathbf{r},\mathbf{r}') \left[ E(J) + \dot{A}_\phi(\rho',z') \right] \tag{1}$$

where $A_\varphi$ is the vector potential of applied magnetic field,

$$Q_{cyl}(\mathbf{r},\mathbf{r}') = f(\rho,\rho',z-z') \tag{2}$$

and $Q^{-1}$ is the reciprocal kernel, which is defined by

$$\int_0^a d\rho' \int_0^b dz' Q^{-1}(\mathbf{r},\mathbf{r}') Q(\mathbf{r}',\mathbf{r}'') = \delta(\mathbf{r}-\mathbf{r}'') \tag{3}$$

with

$$f(\rho,\rho',z-z') = \int_0^\pi \frac{d\phi}{2\pi} \frac{-\rho'\cos\phi}{\left[(z-z')^2 + \rho^2 + \rho'^2 - 2\rho\rho'\cos\phi\right]^{1/2}}$$

$$= \frac{-1}{\pi k}\sqrt{\frac{\rho'}{\rho}}\left[\left(1-\frac{1}{2}k^2\right)K(k^2)-E(k^2)\right] \quad \text{and} \quad k^2 = \frac{4\rho\rho'}{(\rho+\rho')^2+(z-z')^2}$$

Eq. (1) can be easily time integrated by starting with $J(\rho,z,t_0)=J_0$ ($J_0$ is the initial current density distribution in the SD disk) and then by putting $J(\rho,z,t=t+dt)=J(\rho,z,t)+\dot{J}(\rho,z,t)dt$. As soon as the induced current density $J(\rho,z,t)$ is obtained, the vector potential generated by the induced current density $A_J$ can be derived as,

$$A_J(\rho,z) = -\mu_0 \int_0^a d\rho' \int_0^b dz' Q(\mathbf{r},\mathbf{r}')J(\mathbf{r}'). \tag{4}$$

And the radial and axial trapped magnetic field can be written in the form of,

$$B_\rho = -\frac{\partial A_J}{\partial z}, \quad B_z = \frac{1}{\rho}\frac{\partial(\rho A_J)}{\partial \rho} \tag{5}$$

respectively.

These equations should be supplemented by relationships between $J$ and the magnetic field $B$ and the electric field $E$, which depends on the material. In the superconducting state region, where $J < J_c$, $J_c$ is the critical current density, the power-$n$ model is used to describe the nonlinear characteristics of the superconductor [9]:

$$E = E_c \left(\frac{J}{J_c}\right)^n \tag{6a}$$

$n = \sigma+1$ and $\sigma$ is the flux creep exponent, and in the normal state region, where $J \geq J_c$, we use

$$E(J) = E_c \left(\frac{J}{J_c}\right) \tag{6b}$$

Generally the critical current density depends on both the local field $B$ and the temperature $T$. The Kim model is used to describe the flux density dependence of $J_c$ [10]:

$$J_c = J_{c0}\frac{B_0}{|B|+B_0} \tag{7}$$

where $J_{c0}$ is $J_c$ when $B=0$ and $B_0$ is a parameter. We include the temperature dependence of $J_c$ as the following equation [10]:

$$J_{c0} = \alpha\left[1-\left(\frac{T}{T_{c0}}\right)^2\right]^2 \tag{8}$$

where $T_{c0}$ is the critical temperature at $B = 0$ T and $\alpha$ is constant.

## Heat dissipation

When a superconductor is subjected to a non-stationary external magnetic field, the heat generation rate per unit volume is

$$W = \vec{E} \cdot \vec{J} \tag{9}$$

The temperature change due to the heat generation is described by the heat diffusion equation.

$$C\frac{\partial T}{\partial t} - \kappa\nabla^2 T = W \tag{10}$$

here $\kappa$ is the thermal conductivity, and $C$ is the heat capacity per unit volume.

## The levitation force

According to the similar discussion as before [9], the top surface center of the SD approaches the bottom surface center of PM as,

$$s = z_0 - vt \tag{11}$$

where velocity $v$ represents the speed at which the PM approaches the SD, $z_0$ is the initial distance between the top surface of the SD and the bottom surface of the PM.

As the current density $J(\rho,z,t)$ and the radial magnetic field $B_\rho^{PM}$ inside the SD have been derived, the vertical levitation force along the z-axis can be readily obtained as [9],

$$F_z = 2\pi\int_0^a \rho\, d\rho \int_{-b}^{b} dz\, J(\rho,z) B_\rho^{PM}(\rho,z) \tag{12}$$

## Selection of Parameters

The parameters used in our calculation are taken as follows: $T_{c0} = 92$ K [9]; $\mu_0 = 4\pi \times 10^{-7}$; $C = 0.88\times 10^6$ J/m$^3$K [8]; $\kappa = 6$ W·m$^{-1}$·K$^{-1}$ [8]; $E_c = 1\times 10^{-4}$ V/m [10]; $\alpha = 1.9\times 10^9$ A/m$^2$; $v = 2$ mm/s; $B_0 = 0.5$T and the flux creep exponent $\sigma = 4$. All these parameters remain unchanged in the following discussions.

**Results and Discussions**

The experimental results of the levitation force measured in both the FC and ZFC cases are shown in Fig.2. It can be observed that the levitation force in the FC case ($\beta = 0.0025$ T/s) has a much larger value than that in the ZFC case. Taking the distance of 10 mm for example, the levitation force is 45 N and 10 N for the FC and ZFC cases respectively. In the FC magnetization process, after the applied magnetic field is swept down to zero, large induced current remains in the SD, which dominates the levitation force, while the induced current in the SD by PM's approaching is much smaller than that after FC [9]. But no induced current exists in the SD after the ZFC. There is only a small induced current in the SD when the SD approaches the PM. So it is reasonable that the high levitation force is obtained after the FC. The solid curves in Fig.2 represent our calculated levitation forces using Equation (12). The calculation results are in perfect agreement with the experimental data.

How the rate $\beta$ affects the levitation force is also shown in Fig.2. At lower rate the gradient of magnetic field in the edge is smaller, driving force is weaker, and less magnetic flux escape from the SD. Thus the trapped magnetic field at the center of the SD can be higher after the applied magnetic field is swept down to zero and the induced current density in the SD is larger, which is corresponding to higher levitation force.

When the rate $\beta$ keeps constant 0.0025 T/s, the magnitude $B_0$ of the applied magnetic field becomes a crucial factor. Fig.3 shows the levitation force as a function of the distance *s* with different $B_0$. We measure larger levitation forces with a larger applied magnetic field, which is reasonable because a larger current is induced in the SD with a larger applied magnetic field and thus generates a larger levitation force when the PM approaches at a constant velocity $v$.

## Conclusion

We have set up a levitation force measurement system and measured the levitation forces between a permanent magnet and a coaxial superconducting disk after different magnetization processes. The effects of the magnitude of applied magnetic field and the rate on the levitation force are studied in both experiment and theoretical calculation. The calculated results coincide with the experimental data very well, which strongly supports the systematic theoretic framework proposed by Hou *et al.*.

## Acknowledgments

This work was supported by the National Science Foundation of China (NSFC 10174004), the Ministry of Science and Technology of China (Project No. NKBRSF-G1999064602) and the TaiZhao Foundation of Peking University.

Figure captions:

Fig. 1: Experimental setup for the levitation force measurement.

Fig. 2: The measured levitation forces $F_z$ as a function of distance *s* with different $\beta$. The solid curves are calculated results.

Fig. 3: The measured levitation forces $F_z$ as a function of distance *s* with different magnitudes of the applied magnetic field.

Fig. 1

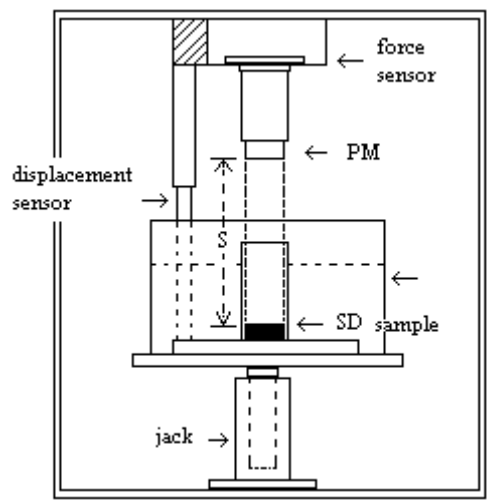

Fig. 2

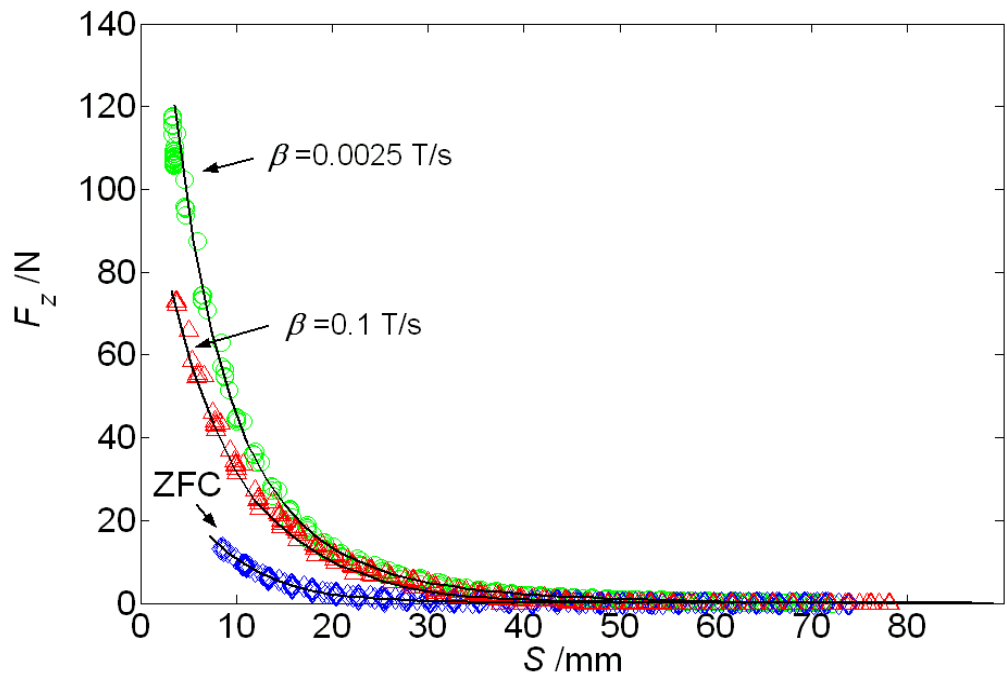

Fig. 3

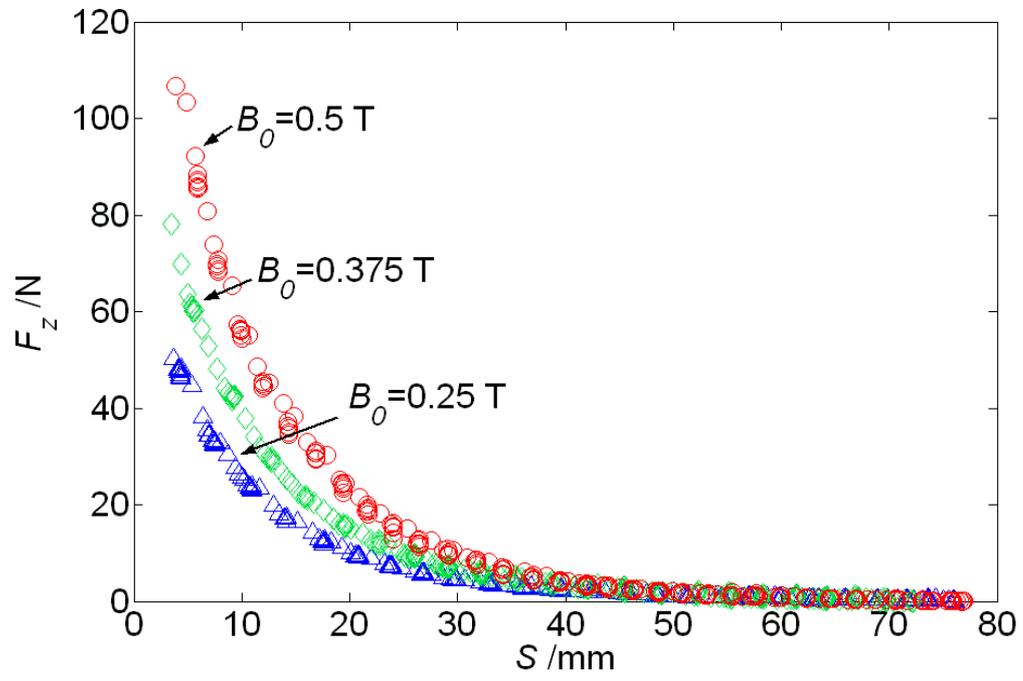